\documentclass[final,3p,times]{elsarticle}
\usepackage{amssymb,bbold}
\usepackage{amsmath}
\usepackage[utf8x]{inputenc}
\usepackage{natbib}
\usepackage{hyperref}

\hypersetup{
  colorlinks=true,linkcolor=blue,citecolor=blue,
  filecolor=blue,urlcolor=blue,breaklinks=true
}
\biboptions{sort&compress}
\journal{arXiv}
\RequirePackage{color}

\begin{document}

\begin{frontmatter}
\title{Influence of rotation on the electronic states, magnetization and persistent current in 1D quantum ring}

\author{Lu\'{i}s Fernando C. Pereira}
\ead{luisfernandofisica@hotmail.com}
\author{M\'{a}rcio M. Cunha}
\ead{marciomc05@gmail.com}
\author{Edilberto O. Silva}
\ead{edilbertoo@gmail.com}
\address{
	Departamento de F\'{i}sica,
	Universidade Federal do Maranh\~{a}o,
	Campus Universit\'{a}rio do Bacanga,
	65085-580 S\~{a}o Lu\'{i}s-MA, Brazil
}

\begin{abstract}
Inertial effects can affects several properties of physical systems. In particular, in the context of quantum mechanics, such effects has been studied in diverse contexts.  In this paper, starting from the Schr\"{o}dinger equation for a rotating frame, we describe the influence of rotation on the energy levels of a quantum particle constrained to a one-dimensional ring in the presence of a uniform magnetic field. We also investigate how the persistent current and the magnetization in the ring are influenced by temperature and rotating effects.
\end{abstract}

\begin{keyword}
Quantum ring \sep
Rotating effects
\sep
Magnetization
\sep
Persistent current
\end{keyword}

\end{frontmatter}

\section{Introduction}

 Quantum Mechanics is one of most well-succeed scientific theories, with a wide range of interesting phenomena and applications. In Condensed Matter Physics, it is an essential tool in describing the behavior of physical systems. In this context, low dimensional systems, for instance, consist of a very fruitful subject for investigation, because of their emergent physics. Such materials have been attracted much attention nowadays.

Since the experimental discovery of graphene \cite{allen2009honeycomb} and posterior increasing interest in carbon structures based-materials \cite{tang2013graphene}, a fast development took place. Several types of structures in the nano and mesoscopic scales can be imagined and synthetized.

In the context of the physics of low dimensions materials, an interesting example of systems are quantum rings \cite{viefers2004quantum,fomin2013physics}. Besides the small size, these structures are fascinating \cite{fuhrer2001energy} and exhibit  several possibilities for investigation, covering since fundamental aspects of the quantum mechanics to applications in condensed matter physics. We can list some of these features. For instance, the effects of the confining quantum particles due to some types of potential can be investigated. Another possibility is the study of magnetic properties such as magnetization in a quantum ring \cite{PhysRevB.70.081301}.  Another aspect that could be explored is the emergence of persistent currents in such systems when in the presence of magnetic flux. In all these cases, temperature plays a fundamental role, since the coherence phase length $L_{\phi}$ increases significantly at low temperatures \cite{PRB.1988.37.6050}.

Several models of quantum rings has been proposed in the literature. A well-know model to study electronic properties of such systems it is due to Tan and Inkson, presented in 1996 \cite{tan1996electron}. In their model, a quantum particle it is constrained to a ring due the presence of a radial confining potential $V(r)$, which $r$ being the radial coordinate in the $x-y$ plane. Posteriorly, they extended the model and have also incorporated some important discussions about the existence of persistent currents and magnetization in such systems \cite{PhysRevB.60.5626}.

After these contributions from Tan and Inkson, other models to describing two dimensional rings also took place. Bulaev and collaborators, for instance, have studied the effect of surface curvature on the persistent currents \cite{PhysRevB.69.195313}.
In Ref. \cite{PhysRevB.72.125348}, it is make a theoretical investigation on persistent currents in distorted quantum rings, taking into account a geometrical potential \cite{PhysRevA.23.1982} to incorporate curvature effects.
In Ref. \cite{PhysRevB.77.205303}, the energy levels of quantum rings of arbitrary shape threaded by a magnetic field are studied.
The study of physical properties of quantum rings also has been done by considering some specific materials. In Ref. \cite{zhang2013aharonov}, for instance, it is considered a graphene ring in the presence of a sinusoidal periodic magnetic flux. Effects of nonzero temperature in the calculation of persistent currents in a zig-zag hexagonal graphene ring are discussed in Ref. \cite{omidi2015temperature}, by employing the tight-binding approach \cite{PhysRevB.40.3979}. More recently, a study about symmetry breaking effects in graphene rings it was reported in Ref. \cite{xu2017symmetry}. The study of two-dimensional quantum rings it is not limited to understanding the electronic states and the persistent currents. Optic absorption in a semiconductor ring, for example, can be controlled by a magnetic field \cite{olendski2014magnetic}. In Ref. \cite{gumber2016optical}, it is investigated how the optical response of a mesoscopic quantum ring it is affected by Rashba spin orbit interaction. In Ref. \cite{olendski2019quantum}, some quantum information measures are calculated for a ring in the presence of magnetic fields.

One-dimensional rings also presents several interesting features. They are a simple model that allow us to investigate relevant physical properties and also applications to other physical systems. An important feature of these devices is that they are very appropriate to study the Aharonov-Bohm effect \cite{PhysRev.115.485,RevModPhys.57.339} for bound states. The Aharonov-Bohm effect establishes that the fundamental quantity in a description of the quantum system is the electromagnetic potential and not the electromagnetic field, which gives a physical significance to the potential vector in quantum mechanics. If we perform a interference experiment in a region where there is a magnetic flux without a correspondent magnetic field, then the wave function acquires a phase. It is also surprising the explanation to the AB effect for bound states \cite{PhysRevA.23.360}: We can imagine a particle confined in a 1D quantum ring, where there is no field but only a magnetic flux \cite{Book.2005.Griffiths}. In this case, the influence of the flux appears modifying the energy levels. In addition, such systems shows a large number of possible applications, and have been explored on different contexts. For example, a study about persistent currents in one-dimensional rings at finite temperature can be accessed in Ref. \cite{PhysRevB.49.8126}. The effects of both spin-orbit and Zeeman interactions on persistent currents in a one dimensional ring it was investigated in Ref. \cite{moskalets2000influence}.

Quantum rings also can be used to perform quantum computing.
In Ref. \cite{PhysRevB.74.125426}, it is addressed the possibility of formation of a flux qubit while in Ref. \cite{szopa2010flux}, a similar discussion it is done to make a flux qubit considering a semiconducting quasi one-dimensional ring.

The study of one-dimensional (hereafter, 1D) rings in the context of relativistic quantum mechanics also has been considered in the literature.
In Ref. \cite{ghosh2014persistent}, for example, it was investigated the behavior of persistent currents on a 1D ring in the presence of scattering potentials. In Ref. \cite{cotuaescu2016relativistic}, it was considered the description of relativistic persistent currents by using the Dirac equation for particles moving in a 1D ring. Another study of one-dimensional quantum rings in the scenario of the Dirac equation can be accessed in Ref. \cite{PhysRevB.88.205401}.

Other aspect of studying 1D quantum rings it is related to the description of thermodynamics properties. In Ref. \cite{PhysRevA.96.013613}, it was studied the thermodynamic behavior of a 1D Bose gas in several aspects, including a ring configuration. In Ref. \cite{oliveira2018thermodynamic}, some thermodynamic properties of AB rings are derived.

All the aspects mentioned above about physical properties of low dimensional systems, particularly quantum rings, involving electromagnetic interactions. On the other hand, another relevant issue in quantum mechanics it is to describe how a system in a non inertial frame can be affected in their quantum mechanical description \cite{PhysRevD.30.368,PhysRevD.15.1448}. Also, it is know that electromagnetic interactions and inertial effects can plays similar influences in physical systems.  In order to clarify this discussion, let us consider some relevant remarks. Let us start making a comparison between the Lorentz force and inertial forces. When an electron moves with linear velocity $\mathbf{v}$ in the presence of an electric field $\mathbf{E}$ and a magnetic field $\mathbf{B}$, it experiences the Lorentz force
\begin{equation}
    \mathbf{F}_{lorentz}=e\mathbf{E}+e(\mathbf{v}\times \mathbf{B})=e[-\mathbf{\nabla}V +\mathbf{v}\times (\mathbf{\nabla}\times \mathbf{A})],
    \label{lorentz_force}
\end{equation}
where $e$ is the electron charge and $V$ and $\mathbf{A}$ are the scalar and the vector electromagnetic potentials, respectively.
Similarly to a particle of mass $m_e$ in a non inertial frame the force is expressed as
\begin{equation}
    \mathbf{F}_{inertial}=2m_e(\mathbf{v}\times \mathbf{\Omega})-m_e\mathbf{\Omega}\times (\mathbf{\Omega}\times \mathbf{r}),
    \label{inertial_forces}
\end{equation}
where $\mathbf{\Omega}$ is the angular frequency and $\mathbf{r}$ is the radius of the orbit. The first term on the right side of the Eq. (\ref{inertial_forces}) refers to the Coriolis force and the second one corresponds to the Centrifugal force. If we consider $\mathbf{\Omega}$ as a constant field, then $\mathbf{\nabla} \cdot  \mathbf{\Omega}=0.$ It means we can write that $\mathbf{\Omega}=\mathbf{\nabla}\times \mathbf{a}$. This way, the rotation can be written in terms of the curl of a vector field $\mathbf{a}$, similarly to the magnetic field $\mathbf{B}$, which can be written in terms of the vector potential $\mathbf{A}$. Also, if we take into account that the centrifugal force it is a conservative force, then it is possible to think in a corresponding scalar potential. After these considerations, we can write the relation
\begin{equation}
\mathbf{F}_{inertial}=m_e[-\mathbf{\nabla}V_{inertial}+2\mathbf{v}\times (\mathbf{\nabla}\times \mathbf{a})].
\label{inertial_forces_2}
\end{equation}
This expression looks like the right side of the Eq. (\ref{lorentz_force}), showing a similarity between electromagnetic fields and rotation.

We can keep forward looking these analogies also considering the context of quantum mechanics. Since the work of Aharonov and Bohm, several analogues of that effect has been presented and investigated \cite{holstein1991variations,bezerra1989gravitational}. It includes a rotational analogue too. More specifically, it is know as the Aharonov-Carmi effect \cite{aharonov1973quantum,aharonov1974quantum}. The main idea of Aharonov- Carmi effect it is the following: Suppose a particle in a rotating ring. The particle it is subjected to the inertial force given by Eq. (\ref{inertial_forces}). Then, in principle, it is possible to apply an electric and a magnetic field in a such way that  the inertial effects and electromagnetic effects cancels one each other. This way, the particle does not feel any forces. However, a quantum phase can arise in a interference experiment performed under these conditions \cite{harris1980review}. Also, it is possible to investigate the effect of rotation on electronic states. In Ref. \cite{shen2005aharonov}, for instance, the Aharonov-Carmi effect it was considered in the context of rotating $C_{60}$ molecules.

Another interesting similarity between rotation and electromagnetic fields can be view from the Barnett effect, where is possible to obtain a magnetization generated by rotation \cite{RevModPhys.7.129} due the coupling between the electron spin and the angular momentum in a rotating sample. Two examples of recent works related to the Barnnet effect can be accessed in Refs. \cite{PhysRevB.92.174424,PhysRevLett.122.177202}.
Other investigations involving rotation in the context of condensed matter physics has been done. In Refs. \cite{fischer2001hall,johnson2000inertial}, are established relations between Hall effect and rotation, for example. A more recent contribution about this subject can be viewed in Ref. \cite{filgueiras2015tuning}. In Ref. \cite{PhysRevB.65.161401}, it was showed that circularly polarized light can spin nanotubes. In Ref. \cite{narendar2011nonlocal}, it was studied wave propagation in a rotating nanotube.
Rotational effects in the context of fullerenes were studied in Refs. \cite{lima2015combined,garcia2017geometric}.

From we have discussed above, we can see how relevant are the investigations regarding quantum rings and rotational effects in quantum mechanics. Then, in this paper, we study a system consisting of a rotating one-dimensional ring in the presence of a uniform magnetic field.
This paper is organized as
follows. In Sec. \ref{sec:model}, we introduce our model and obtain the energy eigenvalues. In Sec. \ref{sec:ele}, we obtain the energy eigenvalues and make the corresponding analysis of the electronic properties. In Sec. \ref{sec:current}, we calculate the temperature-dependent persistent current in the ring and, in Sec. \ref{sec:magnetization}, the temperature-dependent magnetization in the ring. We discuss in detail the effects due to rotation, magnetic flux and temperature on the persistent current and magnetization in the ring. Finally, in the Sec. \ref{sec:conclusions}, we make our conclusions.

\label{sec:introduction}

\section{The model}
\label{sec:model}

\begin{figure}[!htb]
	\centering	
	\includegraphics[scale=0.40]{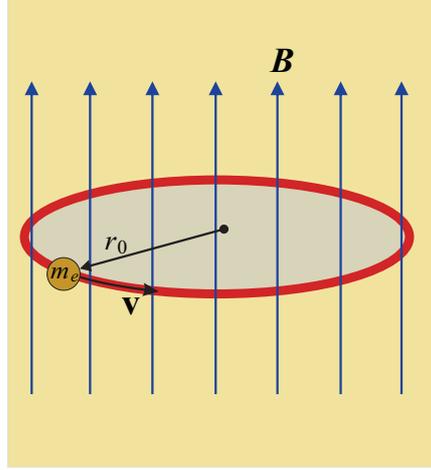}
	\caption{Scheme of an electron executing a circular path of constant radius $r_{0}$ with a velocity $\mathbf{v}$ in the presence of a external magnetic field. The magnetic field $\mathbf{B}$ is uniform.}
	\label{figure_ring}
\end{figure}

In this section, we construct the Schr\"{o}dinger equation that describes the dynamics of a  spinless quantum particle of mass $m_e $ in a rotating frame in the presence of a uniform magnetic field. Following the Ref. \cite{Book_Rotating_Frames}, the nonrelativistic quantum description of a particle in a rotating frame can be done through a Galilei boost with velocity $\mathbf{v}$, given by
\begin{equation}
U=e^{it\mathbf{v}\cdot \mathbf{p}-im_e\mathbf{v}\cdot \mathbf{x}},
\label{boost}
\end{equation}%
connecting two inertial frames $F_{0}$ and $F_{0}^{\prime }$. More explicitally, we have
\begin{equation}
\boldsymbol{x}^{\prime }=\boldsymbol{x}-\mathbf{v}t,\text{ }t^{\prime }=t.
\label{transf}
\end{equation}%
Here, the unprimed coordinates refers to the referential $%
F_{0}$ while the primed coordinates are related to the referential $%
F_{0}^{\prime }$.
The Schr\"{o}dinger equation in the referential $F_{0}^{\prime }$ is written as
\begin{equation}
\left( i\frac{\partial }{\partial t^{\prime }}+\frac{1}{2}m_e v^{2}\right)
\psi \left( \boldsymbol{x}^{\prime },t^{\prime }\right) =\frac{1}{2m_e }%
\left( \mathbf{p}^{\prime }-m_e \mathbf{v}\right) ^{2}\psi \left(
\boldsymbol{x}^{\prime },t^{\prime }\right).   \label{schr}
\end{equation}%
It is interesting to note that this equation can be obtained from the usual Schr\"{o}dinger equation by using the minimal coupling ($\mu=0,1,2,3$)
\begin{equation}
p^{\mu }\rightarrow p^{\mu }-m_e \mathcal{A}^{\mu },  \label{subst}
\end{equation}%
with the following Gauge
\begin{equation}
\mathcal{A}^{\mu }=\left( -\frac{1}{2}v^{2},\mathbf{v}\right) .
\label{fieldA}
\end{equation}%
Now, let us assume that the system described by the Eq. (\ref{schr}) it is in a region where there is a uniform magnetic field in the $z$ direction. Figure \ref{figure_ring} illustrates this model.
An electromagnetic interaction can be included in the Schr\"{o}dinger equation via a minimal coupling, $p^{\mu }\rightarrow p^{\mu }-eA^{\mu }$. In this way,
Eq. (\ref{subst}) can be written as
\begin{equation}
p^{\mu }\rightarrow p^{\mu }-eA^{\mu }-m_e \mathcal{A}^{\mu },
\end{equation}%
with the usual gauge given by
\begin{equation}
A^{\mu }=\left( A^{0},\mathbf{A}\right) ,
\end{equation}%
where
\begin{equation}
\mathbf{A}=\frac{1}{2}Br\,\mathbf{\hat{\varphi}}  \label{poten}
\end{equation}
is the vector potential that generates the uniform magnetic field of Fig. \ref{figure_ring}. Following
 Ref. \cite{Book_Rotating_Frames}, let us assume that $F_{0}^{\prime }$ rotates with a constant angular velocity $\mathbf{\Omega }$ with respect to the frame $F_{0}$.
 The Schr\"{o}dinger equation describing the motion of the system is
\begin{equation}
\frac{1}{2m_{e} }\left[ (\mathbf{p}-e\mathbf{A}-m_{e} \mathbf{\Omega }\times
\mathbf{r})^{2}-\frac{1}{2}m_{e} (\mathbf{\Omega }\times \mathbf{r})^{2}\right]
\psi =E\psi .  \label{schrodingerA}
\end{equation}%
Considering that the particle is constrained to move in a circle of radius $r_0$, Eq. (\ref{schrodingerA}) takes the form
\begin{equation}
\left[ \frac{\hbar ^{2}}{2m_{e}r_{0}^{2}}\left( \frac{\partial }{i\partial
\varphi }-l^{\prime}-\frac{m_{e}\Omega }{\hbar }%
r_{0}^{2}\right) ^{2}-\frac{1}{2}m_{e}\Omega ^{2}r_{0}^{2}\right] \psi
\left( \varphi \right) =E\psi \left( \varphi \right), \label{schrc}
\end{equation}%
with $l^{\prime}=\Phi^{\prime}/\Phi_{0}$, where $\Phi^{\prime}=\pi r_{0}^{2}B$ is the magnetic flux passing through the ring and $\Phi _{0}=h/e$ corresponds to the quantum flux.
The solutions of Eq. (\ref{schrc}) are of the form
\begin{equation}
\psi \left( \varphi \right) =e^{im \varphi }.  \label{sol}
\end{equation}%
The continuity of the wave function $\psi \left( \varphi \right) $ in $%
\varphi =2\pi $ demands that $m$ must be an integer number.
We can solve Eq. (\ref{schrc}) using (\ref{sol}) to find the energy eigenvalues depending of the quantum number $m$ and the physical parameters of the system. Such energies are given explicitly by the expression
\begin{equation}
E_{m} =\frac{\hbar ^{2}}{2m_{e}r_{0}^{2}}\left( m-%
l^{\prime}-\frac{m_{e}\Omega r_{0}^{2}}{\hbar }\right) ^{2}-%
\frac{1}{2}m_{e}\Omega ^{2}r_{0}^{2},
\label{energy_1}
\end{equation}
with $m=0,\pm 1,\pm 2\pm ,\ldots $.
Thus, it generalizes the model of Ref. \cite{Book.2005.Griffiths}, now incorporating rotation effects. This result can be rewritten  as
\begin{equation}
E_{m}=\frac{\hbar ^{2}}{2m_{e} r_0^{2}}\left( m-l^{\prime} \right) ^{2}-\hbar \Omega\left( m-l^{\prime} \right).
\label{energy_2}
\end{equation}%

\section{Electronic properties}
\label{sec:ele}

\begin{figure}[!htb]
	\centering	
	\includegraphics[scale=0.38]{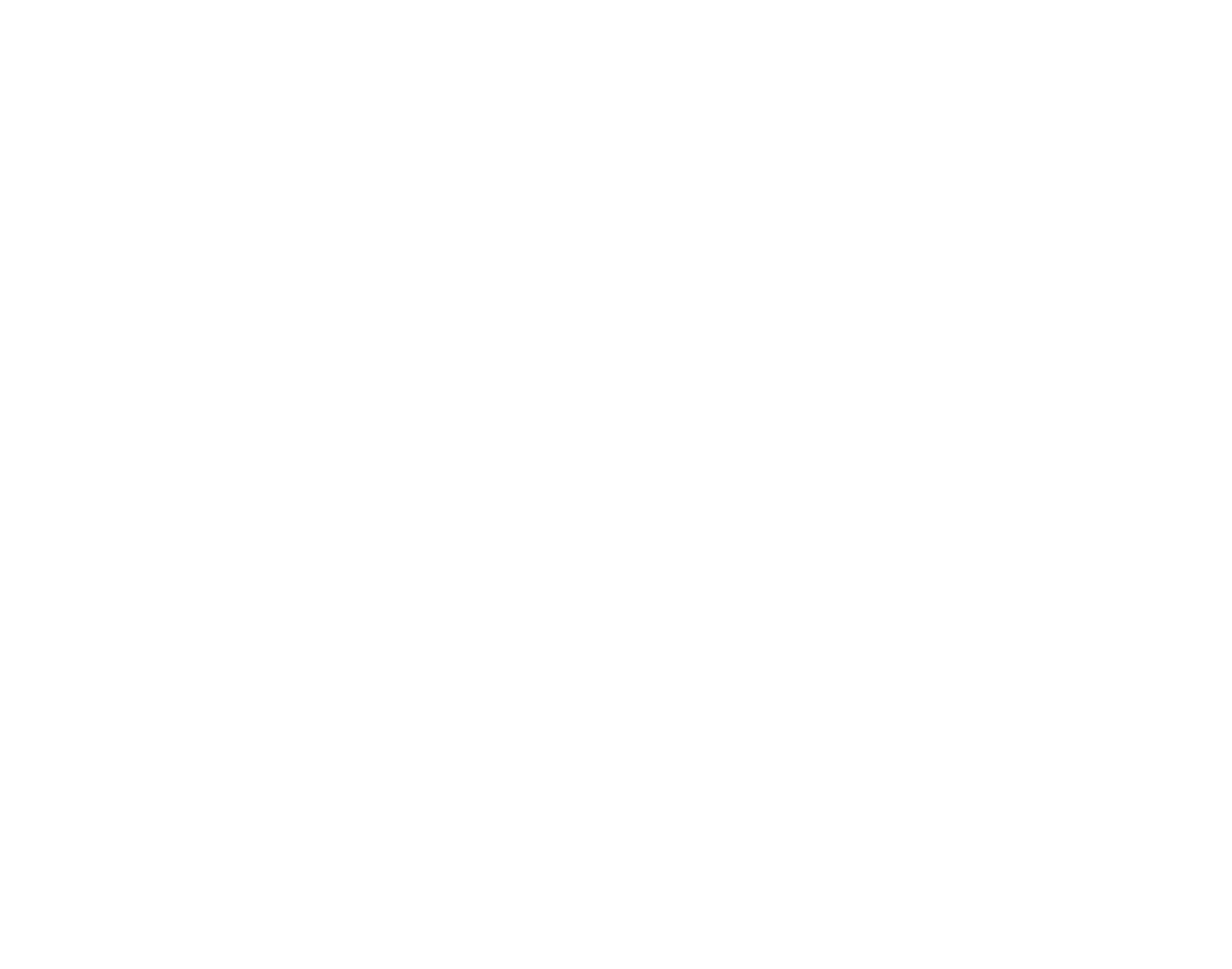}\qquad
	\includegraphics[scale=0.38]{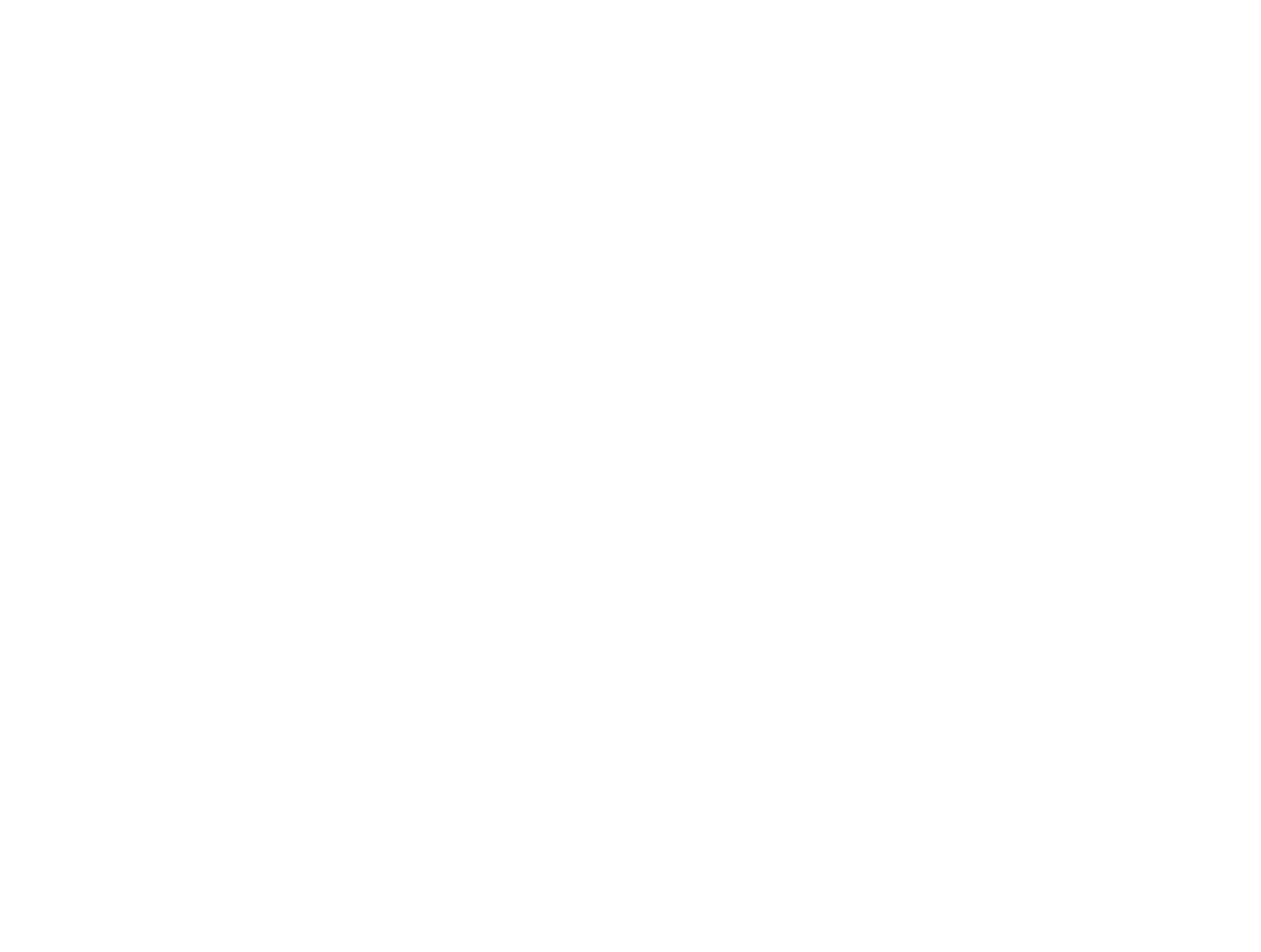}
    \caption{Energy (Eq. (\ref{energy_1})) as a function of $\Phi'/ \Phi_{0}$ for the case without and with rotation.  We consider rings with radius (a) $r_{0}=100$ nm and (b) $r_{0}=400$ nm. Note that for
    $\Phi^{\prime} / \Phi_{0} = 0$ the eigenvalues are doubly degenerate, except for $m=0$. The continuous lines describe the behavior of a single electron.}
	\label{En_r0}
\end{figure}

In this section, we make the analysis of the electronic properties of the system. For a one-dimensional ring in the presence of a magnetic field the energy is modified due the magnetic flux passing through the ring. Considering the energy as function of the magnetic flux, each state $E_m$ describes one parabola with a minimum located at $l'=m$, corresponding to an energy equals to zero. In our model, we have also contributions to the energy due to the rotation.
The rotation introduces changes in the positions of the parabolas on the horizontal axis, corresponding to the minimum
\begin{equation}
l^{\prime}=m-\frac{m_{e} \Omega r_0^2}{\hbar}.
\label{l-min}
\end{equation}
This shows that the relation between the energy minimum and the states are changed due the rotation.
Thus, the parabolas are shifted to the right (left) if the ring is rotating in the clockwise (anticlockwise) direction.
The new energy minimum due the rotation is given by
\begin{equation}
E_{min}=-\frac{1}{2}m_e\Omega^2 r_0^2
\label{E_min}
\end{equation}
and it is independent of the direction of rotation. The results expressed by Eqs. (\ref{l-min}) e (\ref{E_min}) are due the analogy between magnetic field and rotation (Eqs. (\ref{lorentz_force}) and (\ref{inertial_forces}), resulting in the energy eigenvalues given by (\ref{energy_1})).
It is a well-known fact that the energy spectrum of a 1D ring is a periodic function of the magnetic flux, oscillating with a $\Phi_{0}=h/e$ period \cite{Chakraborty2003}. Rotation does not change it.
In order to analyze the behaviour of the system with respect to the magnetic field and rotation in more details, we plot the energy $E_{m}$ as a function of $\Phi^{\prime}/\Phi_0$ for several values of $m$ (Fig. \ref{En_r0}), considering two different ring radius.\vspace{0.2cm}

Figure \ref{En_r0} shows the sketch of the energy as a function of $l'$ for rings with radius $100$ nm (panel (a)) and $400$ nm (panel (b)). We consider  $\Omega=0$ Hz and $\Omega=1.0$ GHz, respectively. Since we are considering an anticlockwise rotation, the parabolas are shifted to the left. We can clearly see that the effects of rotation are more proeminent for the ring of $400$ nm. This characteristic is a manifestation due to the fact that the influence of rotation on the energy depends of the size of the system. Another relevant aspect about these energy states it is related to the degeneracy. We know that the energy of an electron confined in a  1D ring is just $\hbar^2 m^2/2m_e r_0^2$ and the energies are doubly degenerate, except for the $m=0$ state. The introduction of a magnetic field breaks the degeneracy in $m$ (that is, for a given $l'$, the different allowed values of $m$ do not all carry the same energy). This can be checked by analyzing Eqs. (\ref{energy_1}) and (\ref{energy_2}). It is worth to note, however, since the flux is quantized, different combinations of $(m-l')$ can provide the same values of energy. The magnetic flux controls the energy profile and the particle always has the tendency of occupying the lowest level (Fig. \ref{En_r0}).
A particularly interesting case happens when $l'=-m_e \Omega r_0^2/\hbar$, which corresponds to $\Omega=-\omega_c/2$ (being $\omega_c=eB/m_{e}$ the cyclotron frequency). In this case, the rotation cancels the effect of the uniform field on the kinetic term of the energy.
The energy assumes the form
\begin{equation}
E_m(\Omega=-\omega_c/2)=\frac{\hbar ^{2}}{2m_e r_0^{2}}m^2-\frac{1}{2}m_e
r_0^{2}\Omega ^{2}=\frac{\hbar ^{2}}{2m_e r_0^{2}}(m^2-l'^2).
\label{canceling_B}
\end{equation}
Thus, for $\Omega=-\omega_c/2$, differently of the other cases, the influence of rotation on the energy levels corresponds only to an energy shift. This can be justified due the fact that both rotation and magnetic flux contributes to the total angular momentum of the particle. This way, it is possible to tune the field and rotation to modify (or not) the states. In particular, we can define an effective angular momentum given by $j_m=\pm m - (l^{\prime}+\lambda\Omega)$, where $\lambda=m_e r_0^2/\hbar.$

\begin{figure}[!htb]
	\centering	
	\includegraphics[scale=0.7]{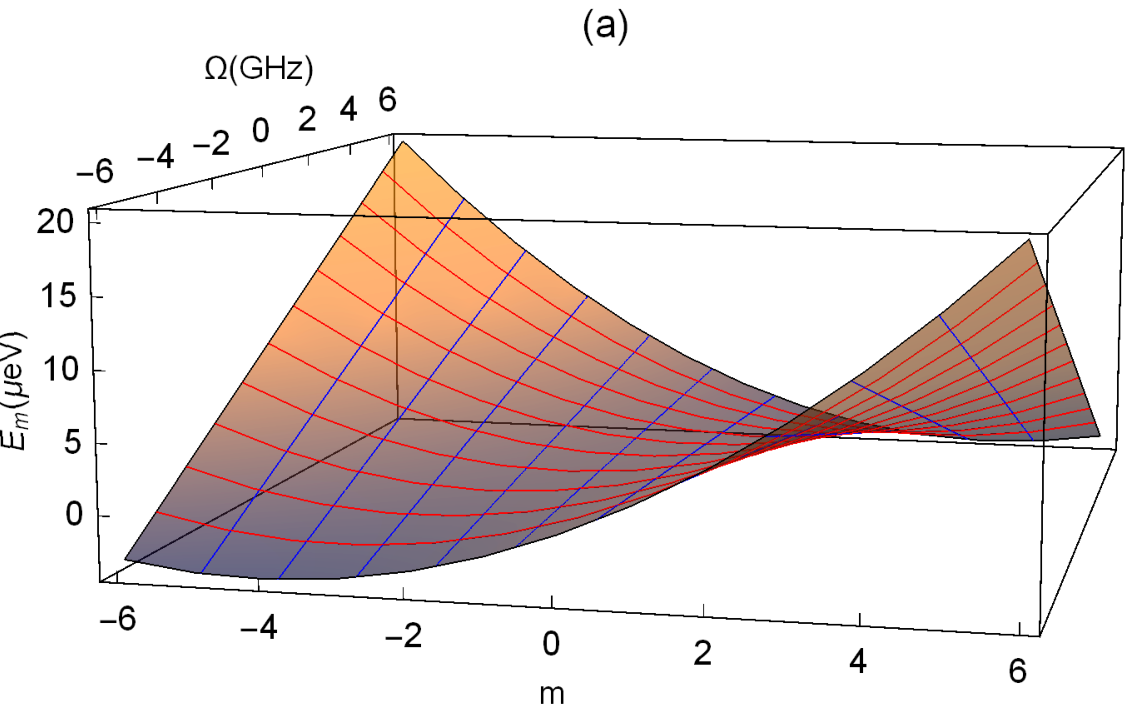}\quad
	\includegraphics[scale=0.65]{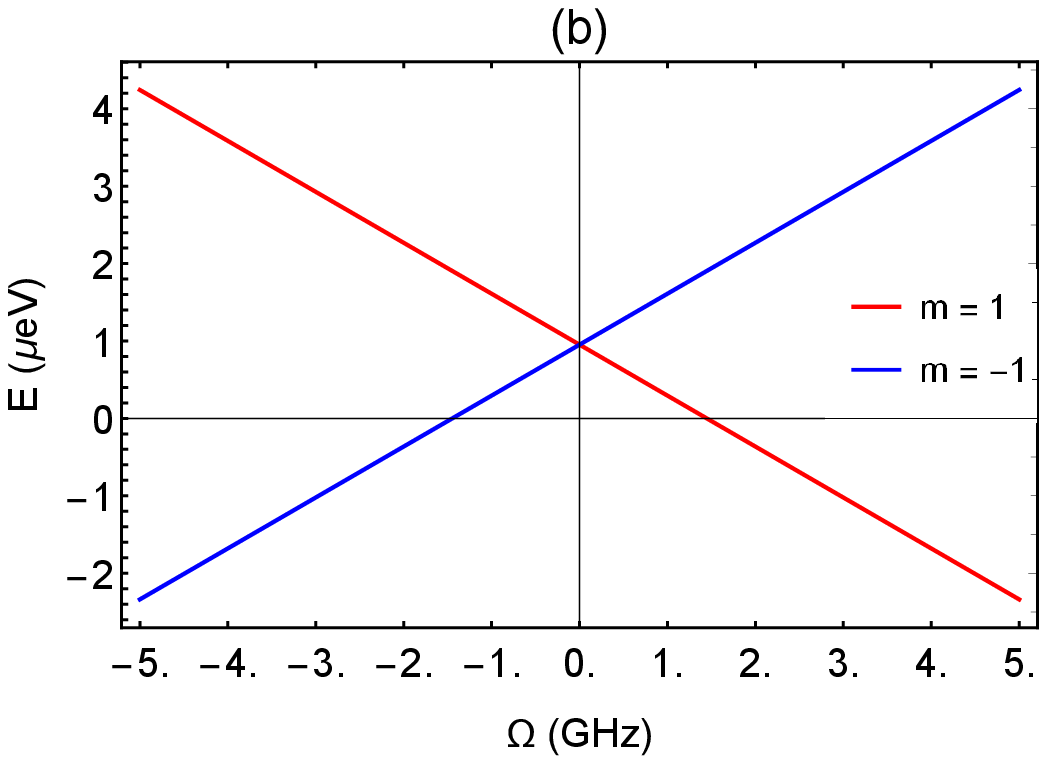}\vspace{0.5cm}
    \caption{In panel (a), the energy profile as a function of $\Omega$ and $m$. In panel (b), the sketch of the energy as a function of $\Omega$, emphasizing only the states with $m = \pm 1$. We consider a ring with radius $r_{0}=400$ nm.}	\label{Em_omega}
\end{figure}

Another relevant point consists in to study only the effect of rotation solely ($l^{\prime}=0$) in the energies. We can think the resultant term $\hbar \Omega m$ in the Eq. (\ref{energy_2}) as a coupling between the rotation and angular momentum of the electron.
This rotation-momentum angular coupling can increase or decrease the energy. For instance, if the ring it is putted to rotate into anticlockwise direction, then a state $-m$ is affected differently than your opposite $m$.
In Fig. \ref{Em_omega}(a), we sketch the energy as a function of $m$ and $\Omega$, considering a ring with radius $r_{0}=400$ nm. As we can observe, the energy eigenvalues have a linear dependence with respect to the rotation. The slope of the function depends of the value of $m$. Also, the slope it is positive (negative) if $m<0$ ($m>0$), as we can see in Fig. \ref{Em_omega}(b), where we show a plot of the energy as a function of $\Omega$, considering the states with $m = \pm 1$.
\begin{figure}[!h]
	\centering	
	\includegraphics[scale=0.7]{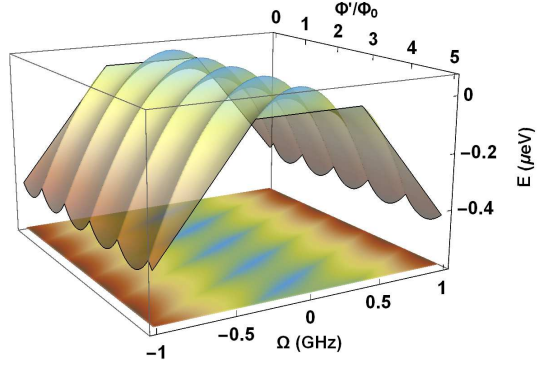}
    \caption{Energy of a single electron as a function of $\Phi'/\Phi_{0}$ and $\Omega$.}
	\label{En3D}
\end{figure}
Before we study the persistent current and magnetization, we can compare the orders of magnitude between the first and the second terms in the Eq. \ref{energy_2}. The first term corresponds to the energy levels of a particle confined in a quantum ring and includes the magnetic flux. The second one refers to the contribution due rotation. Let us consider a ring of $200$ nm,  with $\Omega=10^9$ Hz, $B=0.012$ T and $m=1$. In this case, the first term gives $0.82$ $\mu$eV, while the second one provides $0.61$ $\mu$eV, showing that the inertial contribution can have the same order of magnitude of the first term, depending of the parameters of the system. Figure \ref{En3D} shows the most complete profile of a single electron as a function of the magnetic flux and rotation simultaneously. The states are shifted to the right or left depending on the direction of rotation, while the oscillations are maintained.

\section{Persistent Current}
\label{sec:current}

In this section, we study the temperature-dependent persistent current in a 1D rotating quantum ring. Persistent currents are frequently studied in the context of 1D and 2D quantum rings. Usually, the ring encloses a magnetic flux.
\begin{figure}[!b!]
	\centering	
	\includegraphics[scale=0.7]{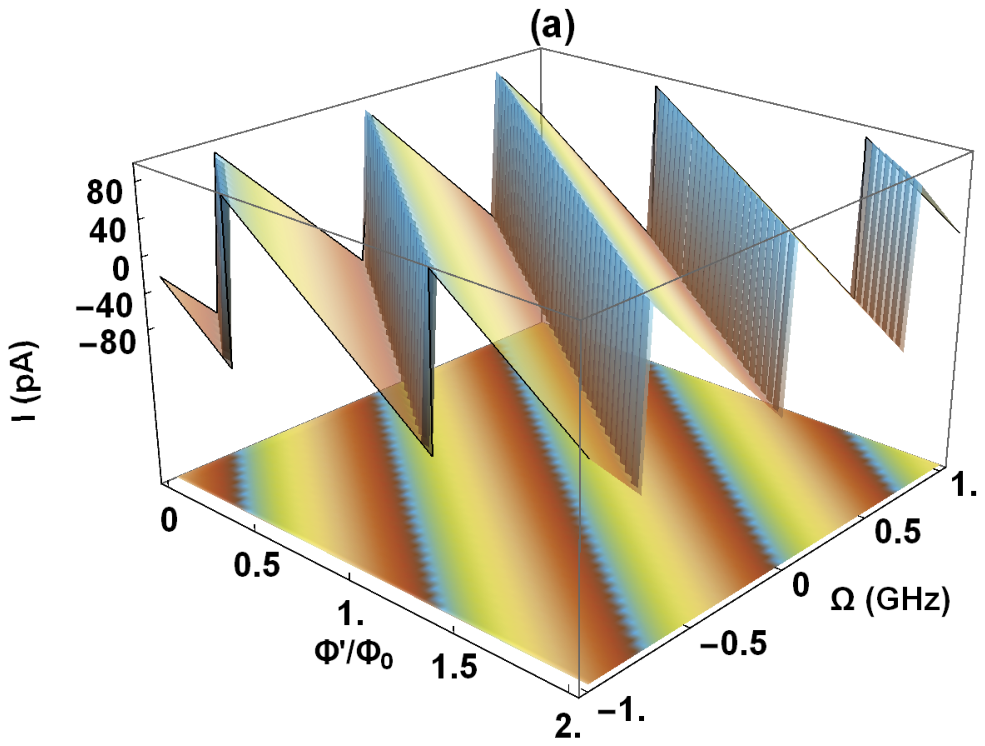}\qquad
\includegraphics[scale=0.7]{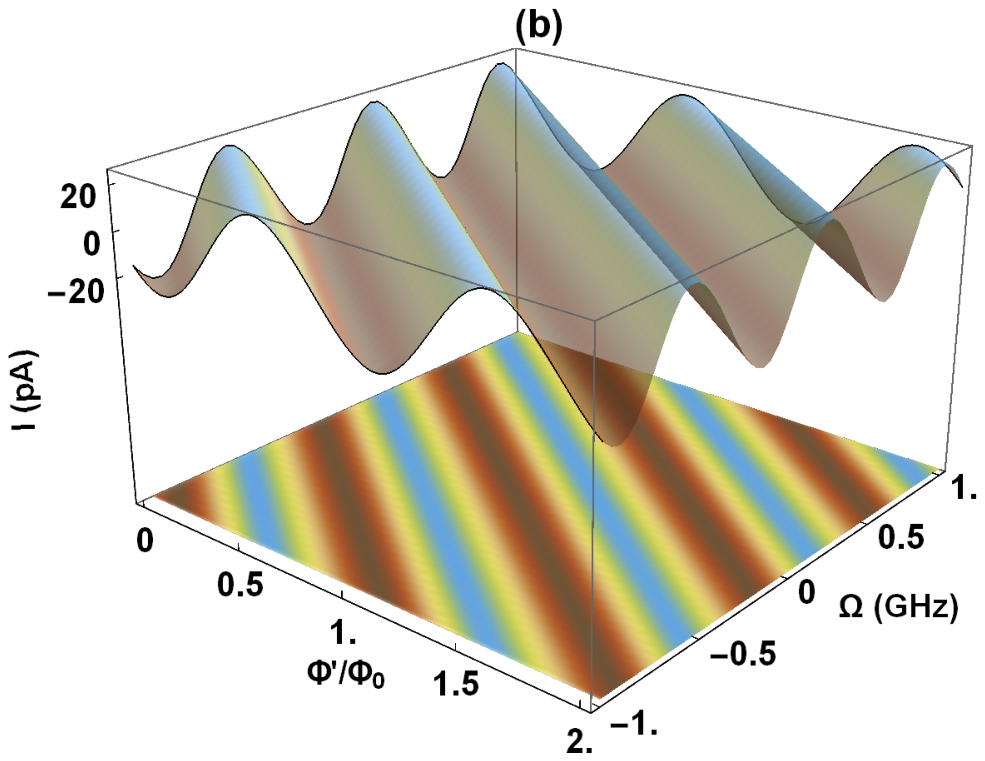}\vspace{0.5cm}
	\includegraphics[scale=0.45]{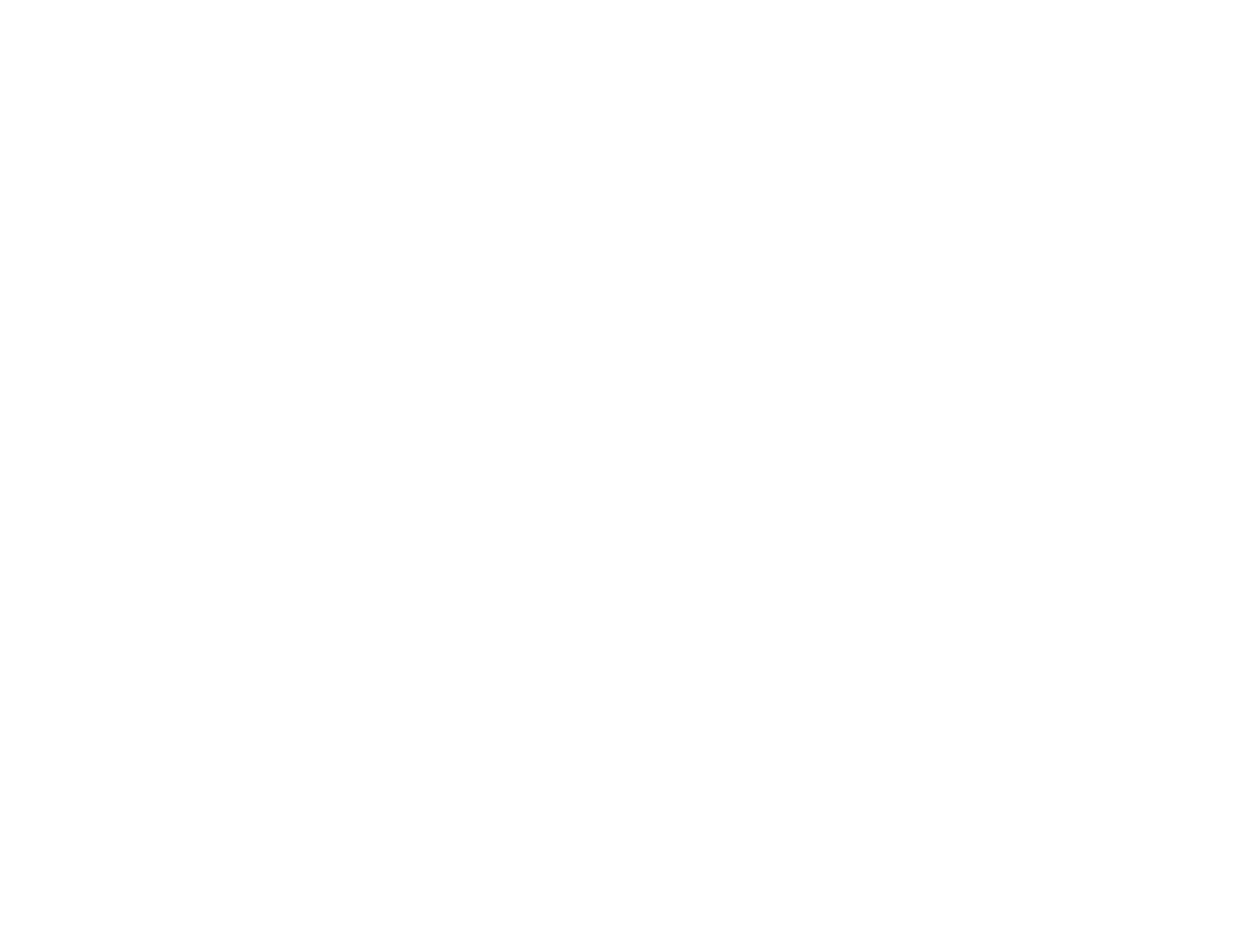}
	\caption{ Persistent currents  as a function $l^{\prime}$ and $\Omega$, for zero temperature (a), and $T=3$ mK (b). In (c), persistent currents as a function of $l^{\prime}$ for $\Omega=-1.0$ GHz (red color line), $\Omega=0$ Hz (green color line) and $\Omega=1.0$ GHz (blue color line). The continuous and dashed lines correspond to T = $0$ K and T = $1.0$ mK, respectively.}
	\label{Cn}
\end{figure}
The persistent current in the ring can be calculated using different approaches. Since the energy explicitly depends on magnetic flux, the easier way to obtain the persistent current is through the Byers-Yang \cite{PRL.7.46.1961} relation
\begin{equation}
I_{m}=-\frac{\partial E_{m}}{\partial \Phi^{\prime} }=-\frac{1}{\Phi _{0}}\frac{%
\partial E_{m}}{\partial l^{\prime}},
\label{Byers-Yang}
\end{equation}%
where $E_{m}$ is given by Eq. (\ref{energy_1}).
We obtain
\begin{equation}
I_{m} =\frac{e\hbar }{2\pi m_{e}r_{0}^{2}}\left( m-%
l^{\prime}-\frac{m_{e}\Omega r_{0}^{2}}{\hbar}\right).  \label{pc}
\end{equation}
The total current it is
\begin{equation}
I=\sum_{m}I_{m}f(E_{m}), \label{Cor}
\end{equation}%
where $f(E_{m})$ is the Fermi distribution function \cite{Landau.2013.Statistical}. We include the Fermi distribution at this point in order to accommodate effects of temperature on the persistent current in the ring \cite{PhysRevB.69.195313}.
When $\Omega=0$ Hz, it is known that the current presents a periodic behaviour with respect to the magnetic flux. If the system it is putted to rotating, a shift on the function $I_m(l')$ is observed. This is intimately related to the shift of the parabolas describing the energy states discussed previously. In Fig. \ref{Cn}, we show the sketch of the persistent current. In Fig. \ref{Cn}(a), we plot both the profile and the corresponding density plot of the current as a function of the magnetic flux $l'$ and rotation $\Omega$ at zero temperature. We can see that the rotation keeps the oscillations, but introduces a linear contribution on the current. In Fig. \ref{Cn}(b), we make a similar plot, but the ring is kept at a temperature $T=3.0$ mK. In Fig. \ref{Cn}(c), we plot the current as a function of the magnetic flux at temperatures $T=0$ K and $T=1.0$ mK, considering $\Omega=0$ Hz, $\Omega=1.0$ GHz and $\Omega=-1.0$ GHz.
As we can see, the temperature tends to smooth out the oscillations as well as the effect of decreasing the persistent current amplitude. Analogously to the case discussed in the previous section, we can also compare the orders of magnitude of the terms in the expression for the persistent current. In this case, we find that the inertial contribution can be comparable with the usual term which depends of $m$ and $l^{\prime}$ when the temperature is zero.
By considering a ring with a radius of  $400$ nm and $\Omega=10^9$ Hz, for example, we can obtain the same order in both terms taking $B=0.01$ T.

\section{Magnetization}
\label{sec:magnetization}

In this section, we study the  temperature-dependent magnetization in the ring. The magnetization $M$ can be evaluated by using the following expression:
\begin{equation}
M=-\frac{\partial U}{\partial B}=-\sum_{m}M_{m}f(E_{m}),  \label{Mag}
\end{equation}
where $M_{m}\equiv -\partial E_{m}/\partial B$ defines the magnetic momentum.
Explicitly, in the present case, we have
\begin{equation}
M_{m} =\frac{e\hbar }{2m_{e}}\left( m-l^{\prime}-\frac{m_{e}\Omega r_{0}^{2}}{\hbar }\right).
\label{Magnetization_eq_2}
\end{equation}%
If the system is not rotating, it is known that the persistent current and the magnetization are related by
\begin{equation}
M_{m}=\pi r_{0}^{2}I_{m}.
\end{equation}
\begin{figure}[!t!]
	\centering
	\includegraphics[scale=0.70]{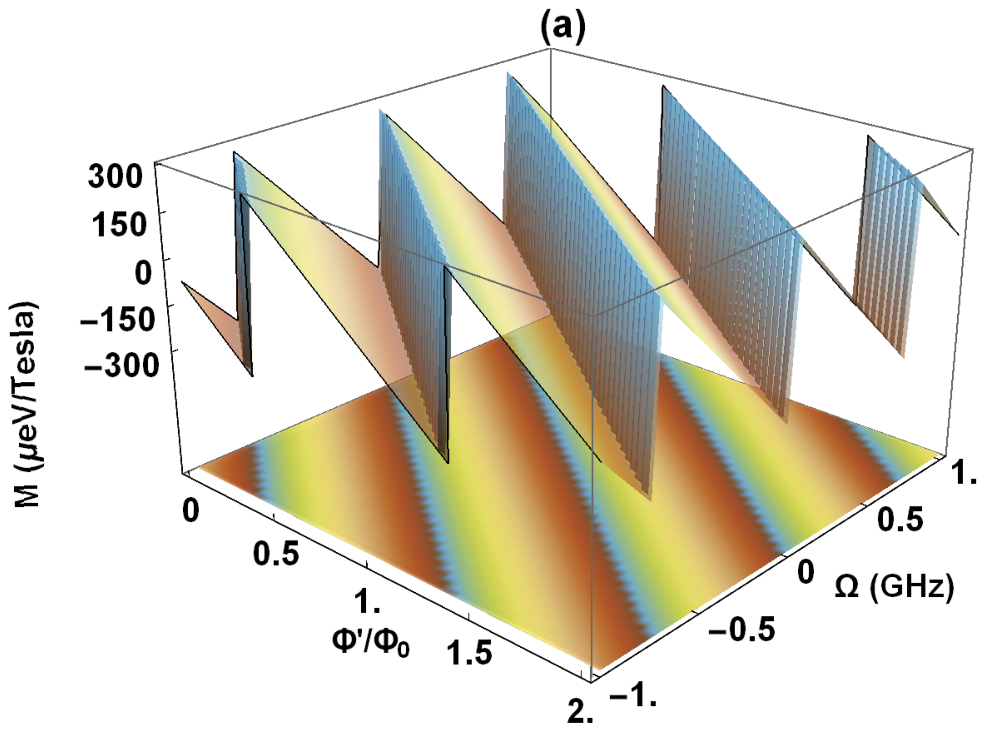}\qquad
\includegraphics[scale=0.70]{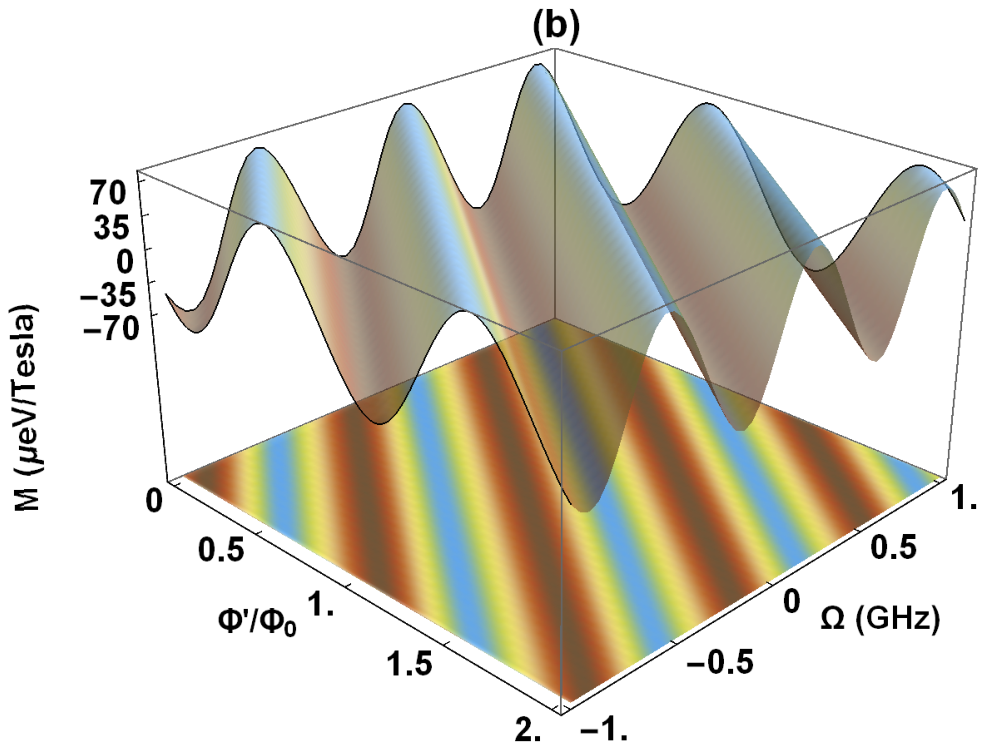}\vspace{0.5cm}
	\includegraphics[scale=0.45]{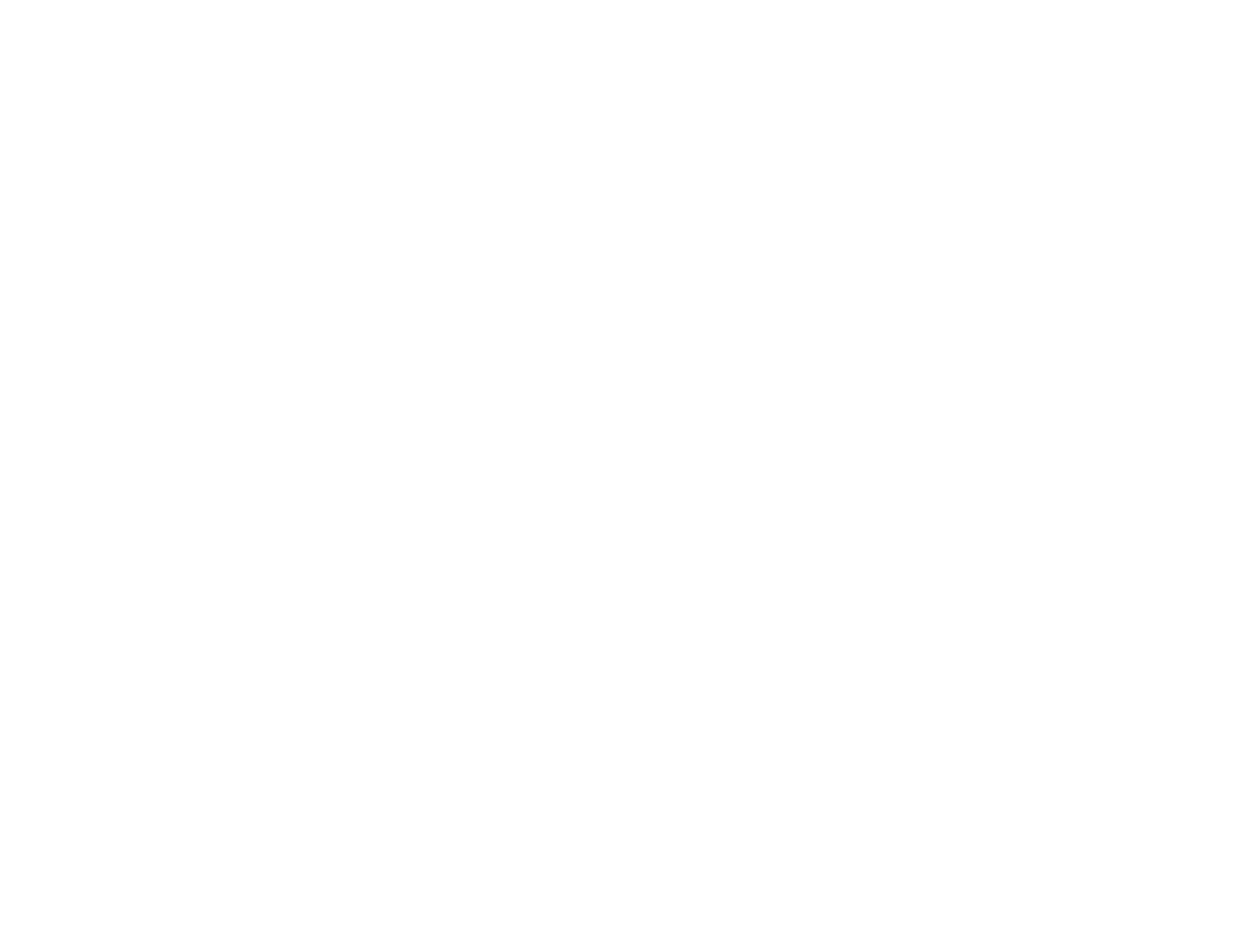}
	\caption{(a) Sketch of the Magnetization as a function of $l^{\prime}$ and $\Omega$ at zero temperature. In panel (b), the ring is kept at temperature $T=3$ mK. In panel (c), the plots of the magnetization as a function of the magnetic flux for $\Omega=-1.0$ GHz, $\Omega=0$ Hz  and $\Omega=1.0$ GHz. The continuous and dashed lines correspond to $T = 0$ K and $T = 1.0$ mK, respectively.}
	\label{Mag_figure}
\end{figure}
This result remains valid when we include inertial effects. In Fig. \ref{Mag_figure}(a), we show a 3D plot of the magnetization as function of $l^{\prime}$ and $\Omega$ at zero temperature. This same magnetization profile at a temperature $T= 3.0$ mK is sketched in Fig. \ref{Mag_figure}(b).

The magnetization as a function of $l^{\prime}$ for some fixed values of $\Omega$ and temperature is sketched in Fig. \ref{Mag_figure}(c).
Clearly, we can observe a similarity between the magnetization profile and the persistent current in Fig. \ref{Cn}, showing that the rotation does not changes the correspondence between current and magnetization. The modifications in the temperature-dependent magnetization profile and at zero temperature due to the rotating effects can have the same order of magnitude that the first term in Eq. (\ref{Magnetization_eq_2}), similarly to we have founded to the energy and persistent current.

\section{Conclusions}
\label{sec:conclusions}

In this paper, we have addressed the problem of a quantum particle constrained to a rotating one-dimensional ring in the presence of an uniform magnetic field. First, we have introduced the Schr\"{o}dinger equation in the case of a rotating frame, and then we specialized to the case of an one-dimensional ring.
Then, we have studied the energy levels of the system. We have discussed the influence due the magnetic flux and rotation. The energy presents an oscillatory behavior with respect to the magnetic flux. Rotation introduces some changes on the energy levels, but maintains the oscillations. Also, we have calculated the persistent current and discussed how both magnetic flux and rotation affects the current. In addition, we have considered temperature effects on the persistent current. Posteriorly, we proceed in a similar way to investigate the magnetization of the system. We have noticed that rotation does not destroy the relationship between current and magnetization. This way, we have investigated how inertial effects affects the electronic states as well persistent current and magnetization. We have founded that depending of the parameters of the system, the inertial effects plays an important contribution on the physical properties studied. More specifically, these effects can have the same order of magnitude of the magnetic effects in some situations. Thus, we hope to contribute in the understanding of inertial effects on quantum systems. Besides the low dimensionality of the system, we can learn a lot from it. In addition, these model can be a tool for future studies involving more complicated systems.

\section*{Acknowledgments}

This work was partially supported by the Brazilian agencies CAPES, CNPq and FAPEMA. EOS acknowledges CNPq Grants 427214/2016-5 and 303774/2016-9, and FAPEMA Grants 01852/14 and 01202/16. MMC acknowledges CAPES Grant 88887.358036/2019-00.

\bibliographystyle{model1a-num-names}

\end{document}